\documentclass[12pt,a4paper]{article}
\pdfoutput=1
\usepackage{epsfig}
\usepackage{comment}
\usepackage{latexsym}
\usepackage{color}
\usepackage{amsmath}
\usepackage{hyperref}
\usepackage[english]{babel}

\newcommand{\mysquare}[0]{\raise-.2ex\hbox{{\Large$\Box$}}}
\def\lsim{\mathrel{\rlap {\raise.5ex\hbox{$ < $}}
{\lower.5ex\hbox{$\sim$}}}}
\def\gsim{\mathrel{\rlap {\raise.5ex\hbox{$ > $}}
{\lower.5ex\hbox{$\sim$}}}} \topmargin -1.5cm \textheight=22.5cm \textwidth=16.5cm
\catcode`\@=11
\newcount\hour
\newcount\minute
\newtoks\amorpm
\hour=\time\divide\hour by60 \minute=\time{\multiply\hour by60 \global\advance\minute by-\hour}
\edef\standardtime{{\ifnum\hour<12 \global\amorpm={am}%
        \else\global\amorpm={pm}\advance\hour by-12 \fi
        \ifnum\hour=0 \hour=12 \fi
        \number\hour:\ifnum\minute<10 0\fi\number\minute\the\amorpm}}
\edef\militarytime{\number\hour:\ifnum\minute<10 0\fi\number\minute}
\def\draftlabel#1{{\@bsphack\if@filesw {\let\thepage\relax
   \xdef\@gtempa{\write\@auxout{\string
      \newlabel{#1}{{\@currentlabel}{\thepage}}}}}\@gtempa
   \if@nobreak \ifvmode\nobreak\fi\fi\fi\@esphack}
        \gdef\@eqnlabel{#1}}
\def\@eqnlabel{}
\def\@vacuum{}
\def\draftmarginnote#1{\marginpar{\raggedright\scriptsize\tt#1}}
\def\draft{\oddsidemargin -.2truein
        \def\@oddfoot{\sl preliminary draft \hfil
        \rm\thepage\hfil\sl\today\quad\militarytime}
        \let\@evenfoot\@oddfoot \overfullrule 3pt
        \let\label=\draftlabel
        \let\marginnote=\draftmarginnote
   \def\@eqnnum{(\theequation)\rlap{\k

 ern\marginparsep\tt\@eqnlabel}%
\global\let\@eqnlabel\@vacuum}  }

\newcommand{\be}[0]{\begin{equation}}
\newcommand{\ee}[0]{\end{equation}}
\newcommand{\ba}[0]{\begin{eqnarray}}
\newcommand{\ea}[0]{\end{eqnarray}}

\def\bs{\begin{subequations}}
\def\es{\end{subequations}}

\def\thebibliography#1{%
\vskip 0.5cm \centerline{\bf \Large References}
\list{%
[\arabic{enumi}]}{\settowidth\labelwidth{[#1]} \leftmargin\labelwidth \advance\leftmargin\labelsep
\usecounter{enumi}}
\def\newblock{\hskip .11em plus .33em minus .07em}
\sloppy\clubpenalty4000\widowpenalty4000 \sfcode`\.=1000\relax}

\renewcommand{\theequation}{\arabic{section}.\arabic{equation}}

\renewcommand{\section}{\setcounter{equation}{0}\@startsection
{section}{1}{0mm}{-\baselineskip}{0.5\baselineskip} {\normalfont\Large\bfseries}}

\renewcommand{\subsection}{\@startsection
{subsection}{2}{0mm}{-\baselineskip}{0.5\baselineskip} {\normalfont\large\bfseries}}

\renewcommand{\subsubsection}{\@startsection
{subsubsection}{3}{0mm}{-\baselineskip}{0.5\baselineskip} {\normalfont\normalsize\slshape}}

\usepackage{amssymb,amsfonts}
\usepackage{graphicx}
\usepackage{cite}

\newcommand{\bea}{\begin{eqnarray}}
\newcommand{\eea}{\end{eqnarray}}
\newcommand{\dis}{\displaystyle}

\newcommand{\ZZ}{\mathbb{Z}}
\newcommand{\Z}{{\cal Z}}

\renewcommand{\O}{{\cal O}}

\newcommand{\abs}{|}
\newcommand{\Tr}{\textrm{Tr}\, }

\newcommand{\ie}{{\em i.e. }}
\newcommand{\where}{\mbox{where}}

\newcommand{\when}{\mbox{when}}
\renewcommand{\and}{\mbox{and}}

\newcommand{\F}{{\cal F}}

\renewcommand{\S}{{\cal S}}

\renewcommand{\b}{\bar}

\renewcommand{\t}{\tilde}

\begin{document}
\begin{titlepage}
\begin{flushright}
CPHT--PC047.061 \\
June 2011
\end{flushright}

\vspace{2mm}

\begin{centering}

{\bf \Large NON-SINGULAR SUPERSTRING \\
\vspace{3mm}
 COSMOLOGY IN TWO DIMENSIONS}

\vspace{10mm}
 {\Large Herv\'e Partouche}

\vspace{4mm}

Centre de Physique Th\'eorique, Ecole Polytechnique$^\dag$,
\\
F--91128 Palaiseau cedex, France\\
{\em herve.partouche@cpht.polytechnique.fr}

\vspace{8mm}

{\it Based on a talk given at the ``10th Hellenic School and Workshops on Elementary Particle Physics and Gravity'', Corfu, Greece, September 4 - 18, 2010.}

\vspace{10mm}

{\bf\Large Abstract}

\end{centering}
\vspace{4mm}

\begin{quote}
We review a recently proposed approach to construct superstring cosmological evolutions, which are free of Hagedorn instabilities and initial singularities. We illustrate  these ideas in hybrid models in two dimensions. 
\end{quote}

\vspace{3pt} \vfill \hrule width 6.7cm \vskip.1mm{\small \small \small
\noindent
$^\dag$ Unit{\'e} mixte du CNRS et de l'Ecole Polytechnique,
UMR 7644.}

\end{titlepage}
\newpage
\setcounter{footnote}{0}
\renewcommand{\thefootnote}{\arabic{footnote}}
 \setlength{\baselineskip}{.7cm} \setlength{\parskip}{.2cm}

\setcounter{section}{0}


\section{Introduction}

Large classes of two-dimensional conformal field theories can be used to define superstring backgrounds in Minkowski or anti-de Sitter spaces. Since these vacua describe Universes which are static at tree level, one may wonder under which circumstances quantum effects may correct this statement and lead to cosmological evolutions. Clearly, supersymmetry must be broken in order to violate time-translation invariance. Since we are mostly interested in theories with small cosmological constants (compared to the string scale $M_s$) at the quantum level, it is natural to focus on flat (Minkowski) tree-level backgrounds where supersymmetry is {\em spontaneously} broken, in order to not have quantum corrections of order $M_s$. In such theories, the scale $M$ of supersymmetry breaking is a field, whose vacuum expectation value is a flat direction of the vanishing classical potential. As a result, there is no preferred scale for the quantum corrections. 

A physically relevant and general way to construct such ``no-scale models" from supersymmetric vacua is to switch on finite temperature. In this case, the scale of spontaneous supersymmetry breaking is the temperature $T$, while the effective potential is nothing but the free energy. Already at the one-loop level, the latter induces a backreaction on the originally static background which can enter in quasi-static evolution. 

For a system of particles with Hamiltonian $H$,  the canonical partition function $\Z=\Tr e^{-\beta H}$ can be computed in quantum field theory with a path integral in compact Euclidean time of period $\beta=2\pi R_0$, the inverse temperature. In the case of a supersymmetric spectrum, the free energy $F=-(\ln \Z)/\beta$ is finite in the UV and, in the perfect gas approximation, its expression at one-loop can be coupled to the classical action of the system. However, due to the presence of the supergravity multiplet, a rigorous derivation of this procedure in quantum field theory is lacking. However, the previous 1-loop corrected supergravity action can be obtained as the low energy effective description of superstring models at finite temperature. In this case, $Z=\ln \Z$ is the genus-1 vacuum-to-vacuum amplitude evaluated in Euclidean time compactified on $S^1(R_0)$.  

The above statements have however some limitation. When the temperature increases, the number of string modes which can be thermalized grows exponentially and implies $Z$ to diverge above some Hagedorn temperature $T_H$ of order the string scale. This breakdown of the canonical ensemble approach is believed to signal a phase transition at the maximal temperature $T_H$ towards another (possibly stringy) thermal system. However, what may look as a difficulty string theory has to face may provide a possibility to overcome the initial singularity problem, where both curvature and temperature are infinite. The aim of the present note is to summarize a recently proposed approach to construct cosmological evolutions free of Hagedorn and initial singularities \cite{1,2}.   


\section{Hagedorn singularity free models}
\label{2}

To be specific, we consider the type IIA and type IIB superstrings toroidally compactified on the Euclidean background $S^1(R_0)\times T^{D-1}\times T^{9-D}\times S^1(R_9)$, where  $T^{D-1}$ of volume $V$ regularizes the infinite size of the external space, while $T^{9-D}\times S^1(R_9)$ stands for a factorized internal space. The space-time supersymmetries generated by the right-moving sector are spontaneously broken by coupling the $\Gamma_{(1,1)}$ lattice of zero modes associated to $S^1(R_9)$ to $\b a$, the right-moving Ramond-Ramond (RR) charge. The remaining left-moving supersymmetries can be spontaneously broken by implementing finite temperature. This is done by coupling the Euclidean time $\Gamma_{(1,1)}$ lattice to the fermion number $a+\b a$. In type IIA ($\lambda=0$) and type IIB ($\lambda=1$), the partition function takes the form,
\begin{eqnarray}
\label{Ztherm}
Z\!\!\!\!&=&\!\!\!\!\dis{V\over (2\pi)^{D-1}}  \int_\F {d^2\tau\over 2\tau_2^{D+1\over 2}} \, \dis {\Gamma_{(9-D,9-D)}\over \eta ^{8}\b \eta^{8}}\, {1\over 2}\sum_{a,b}(-)^{a+b+\lambda ab}\, {\theta[^a_b]^4\over \eta^4} \, {1\over 2}\sum_{\b a,\b b}(-)^{\b a+\b b+\b a\b b}\, {\b \theta[^{\b a}_{\b b}]^4\over\b \eta^4}\\
&&\!\!\!\!\dis {R_0\over \sqrt{\tau_2}}\!\sum_{n_0,\t m_0}\!\!e^{-{\pi R_0^2\over \tau_2}\abs n_0\tau+\t m_0\abs^2} \!(-)^{(a+\b a)\t m_0+(b+\b b)n_0} {R_9\over \sqrt{\tau_2}}\!\sum_{n_9,\t m_9}\!\!e^{-{\pi R_9^2\over \tau_2}\abs n_9\tau+\t m_9\abs^2}\! (-)^{\b a\t m_9+\b b n_9+\t m_9 n_9} .\nonumber
\end{eqnarray}

An alternative way to break the left-moving supersymmetries is to couple the Euclidean time lattice to the left-moving RR charge only. This amounts to changing 
\be
(-)^{(a+\b a)\t m_0+(b+\b b)n_0}\quad \longrightarrow \quad  (-)^{a\t m_0+bn_0+\b m_0n_0}
\ee
in Eq. (\ref{Ztherm}). In both cases, the one-loop correction to the vacuum energy is non-trivial and backreacts on the classically static background. Our aim is to make the comparison between the two choices \cite{1}. 

For either couplings, the partition function can be expressed in terms of $SO(8)$ affine characters $V_8=S_8=C_8$ and $O_8$, by  Poisson resumming on the indices $\t m_0$ and $\t m_9$. Due to the GSO projections which are reversed when the winding numbers $n_0$ or $n_9$ are odd, there exist $V_8\b O_8$ sectors whose lightest states of mass $\abs R_9-1/(2R_9)\abs$ enhance $U(1)_R\to SU(2)_R$ when $R_9=R_c:=1/\sqrt{2}$. Since this point is a minimum of $-Z$, it is consistent to suppose from now on $R_9$ to be stabilized at $R_c$. There are also $O_8\b O_8$ sectors. In the thermal models, the associated lightest modes have squared masses $m_{O\b O}^2=R_0^2-2$. They become tachyonic for $R_0<R_H:=\sqrt{2}$ and imply $Z$ to be ill-defined above the Hagedorn temperature $T_H=1/(2\pi R_H)$. This pathology may be cured by allowing the tachyons to condense \ie acquire vacuum expectation values when they become massless. Below $T_H$, one can make contact between the genus-0 amplitude $Z$ and the canonical ensemble partition function  $\Z=\Tr e^{-\beta H}$ by ``unfolding'' the fundamental domain $\F$ in Eq. (\ref{Ztherm}). What is meant by this is that integrating over $\F$ the discrete sum over $n_0$ is equivalent to integrating  over the upper half strip $\S_+$ $\left(-{1\over 2}<\tau_1<{1\over 2},\,  \tau_2>0\right)$ and setting $n_0=0$. It is then straightforward to integrate over $\tau_1$ to implement level matching and identify  $Z$ with $\ln \Z$, when the latter is expressed in first quantized formalism as an integral over the Schwinger parameter $\tau_2>0$. Since $R_0$ is constrained to be larger than $R_H$, there is no allowed T-duality on $S^1(R_0)$ which may relate the IIA and IIB models.   

On the contrary, the lightest states in the $O_8\b O_8$ sector of the second class of models have masses $m_{O\b O}^2=[R_0-1/(2 R_0)]^2$ and are never tachyonic. $R_0>0$ is allowed to take arbitrary values and the IIA and IIB models are actually identified, since 
\be
Z(\lambda,R_0)=Z\Big(1-\lambda,{1\over 2R_0}\Big).
\ee
 The unfolding procedure of the fundamental domain obtained by exchanging the order in which integration over $\tau$ and summation over $n_0$ are performed is valid when absolute convergence of the integrand is satisfied. This is the case when $R_0>R_c$ and leads (say for $\lambda=1$)
\begin{align}
\label{Znontach}
\dis Z={V\over (2\pi)^{D-1}} \int_{\S_+}&\dis {d^2\tau\over 2\tau_2^{D+1\over 2}} \, {\Gamma_{(9-D,9-D)}\over \eta^8\b \eta^8}\, \sum_{m_0}\sum_{m_9,l_9}\!\left.\left( \Gamma_{m_0,0}V_8-\Gamma_{m_0+{1\over 2},0}S_8\right)\!\right\abs_{R_0}\nonumber\\
&\dis \!\left.\left( \Gamma_{m_9,2l_9}\b V_8-\Gamma_{m_9+{1\over 2},2l_9}\b S_8+ \Gamma_{m_9+{1\over 2},2l_9+1}\b O_8-\Gamma_{m_9,2l_9+1}\b C_8\right)\!\right\abs_{R_9},
\end{align}
where $\Gamma_{\alpha, \delta}(R):=q^{[(\alpha/R)^2+(\delta R)^2]/4}$. For $R_0<R_c$, an analogue expression is obtained by T-duality: $R_0\to 1/(2R_0)$, $S_8\to C_8$. In Eq. (\ref{Znontach}), the fermions (bosons) in the sectors $V_8\b S_8$ and $V_8\b C_8$ ($S_8\b S_8$ and $S_8\b C_8$) have integer ($\mbox{integer}+{1\over 2}$) moments along $S^1(R_0)$ and are thus not thermalized as is usually the case. This is clear since $(-)^a=(-)^{(a+\b a)} (-)^{\b a}$ differs from the standard thermal boundary conditions along $S^1(R_0)$ in the right-moving RR sector ($\b a=1$). Actually, the amplitude $Z$ can be written as \cite{1} 
\be
 \label{dressed}
 Z=\ln \Tr \Big[(-)^{\b a}\, e^{-\beta H} \Big]\quad \where \quad \beta = \left\{\begin{array}{ll}2\pi R_0&\dis \!\when \:\; R_0>R_c\\ \dis 2\pi/(2R_0)&\dis \!\when \;\; R_0<R_c\end{array} \right.\!\!\!,
 \ee
which shows that the tachyon free model can be interpreted as a canonical ensemble deformed by the insertion of the operator $(-)^{\b a}$ in the trace over the multi-particle states. The system admits a ``momentum phase" ($R_0>R_c$) connected to a ``winding phase" ($R_0<R_c$), with space-time fermions in distinct representations (as follows from the exchange $S_8\leftrightarrow C_8$). At the fermionic point ($R_0=R_c$), additional massless states occur, whose crucial role is to trigger the phase transition, as will be illustrated in the next section. They are the would-be tachyons we already mentioned in the $O_8\b O_8$ sector, together with modes in the $O_8\b V_8$ sector which enhance $U(1)_L\to SU(2)_L$. Moreover, it is important to stress that the masses of the non-thermalized states (the sectors $\b a=1$) read from Eq. (\ref{Znontach}) are larger than $1/R_9$ or $R_9$, which are of order the string scale since $R_9=R_c$. As a result, the multi-particle states which contain such excitations are exponentially suppressed by Boltzman's factor as soon as $\beta$ is larger than $\beta_c=2\pi R_c$. This justifies the fact that {\em the conventional thermal model defined in Eq. (\ref{Ztherm}) and the tachyon free one cannot be distinguished when the temperature is below the string scale. The two constructions differ at ultra high temperature only, where the Hagedorn singularity of the former is replaced by a thermal ``winding $\to$ momentum" phase transition.} 

To conclude this section, we signal that the dressing of the canonical ensemble partition function defined in Eq. (\ref{dressed}) can be interpreted as a discrete deformation arising for specific background fields $G_{09}=2B_{09}=1$, when  $\ln\Z$ is evaluated  as a Euclidean path integral. These parameters are not dynamical, as they are Wilson lines for the gauge bosons $G_{\mu 9}$ and $B_{\mu 9}$ along the temporal direction $S^1(R_0)$ \cite{1}.


\section{Example of non-singularity cosmology: The hybrid models}

The phase transition announced in the previous section can be described dynamically in particularly simple models in two dimensions \cite{2}. The generalization in arbitrary dimension can be found in \cite{KPT}. The so-called hybrid A and hybrid B models ($\lambda=0,1$) are superstrings theories in Euclidean space-time $S^1(R_0)\times S^1$, with partition functions 
\be
\label{Zhybrid}
\begin{array}{ll}
\dis Z={V\over 2\pi}  \int_\F {d^2\tau\over 2\tau_2^{3/2}} &\!\!\!\!\!  \dis {\Gamma_{(8,0)}\over \eta^8} {1\over 2}\sum_{a,b}(-)^{a+b+\lambda ab}\, {\theta[^a_b]^4\over  \eta^4} \, (\b V_{24}-\b S_{24}) \\
&\!\!\!\!\! \dis {R_0\over \sqrt{\tau_2}}\sum_{n_0,\t m_0}e^{-{\pi R_0^2\over \tau_2}\abs n_0\tau+\t m_0\abs^2} (-)^{a\t m_0+bn_0\t m_0 n_0},
\end{array}
\ee
where $V$ is the perimeter of the spatial $S^1$ which regularizes the infinite external space. The coupling of the temporal zero modes  to the left-moving RR charge $a$ breaks spontaneously all left-moving supersymmetries. On the same side, the internal CFT is based on the $E_8$ root lattice $\Gamma_{(8,0)}$, while the right-moving sector is particularly simple since $\b V_{24}-\b S_{24}\equiv 24$. This remarkable identity between the $SO(24)$ affine characters signals an exact degeneracy between the {\em massive} modes of the right-moving NS and R sectors. On the contrary, 24 {\em massless} NS states in $\b V_{24}$ have no counterparts in $\b S_{24}=2^{11}\b q+\O(\b q^2)$. This is an example of MSDS structure \cite{MSDS}, which implies all right-moving supersymmetries are broken at the string scale. The right-moving CFT corresponds to an $SU(2)_R^8$ enhanced symmetry point of the internal moduli space where $R_2,\dots,R_9$ are stabilized at $R_c$. 

The properties of the hybrid models are similar to those described for the tachyon free models of section \ref{2}.  In the hybrid B case ($\lambda=1$), the unfolding procedure of the fundamental domain $\F$ leads for $R_0>R_c$
\be
\label{ZhybridS}
\dis Z={V\over 2\pi}\dis \int_{\S_+} {d^2\tau\over 2\tau_2^{3/2}} \, {\Gamma_{(8,0)}\over \eta ^8}\, \sum_{m_0}\!\left.\left( \Gamma_{m_0,0}V_8-\Gamma_{m_0+{1\over 2},0}S_8\right)\right\abs_{R_0}(\b V_{24}-\b S_{24}),
\ee
while a similar expression valid for $R_0<R_c$ is obtained by thermal duality: $R_0\to 1/(2R_0)$, $S_8\to C_8$. Since the analogous results in the hybrid A case are simply obtained by changing $R_0\to 1/(2R_0)$, the hybrid A and B theories are identified. Actually, Eq. (\ref{ZhybridS}) and its counterpart for $R_0<R_c$ can be written as in Eq. (\ref{dressed}), which shows that a thermal momentum phase is connected to a thermal winding phase. This can be made even more explicit by computing exactly the partition function $Z$. Due to the exact cancellation of the massive states in the right-moving sector, level-matching implies $Z$ has non-trivial contributions from massless states only:
\be
{Z\over V} =24 \left\{ \begin{array}{ll}\dis 1/R_0&\dis  \mbox{ for } \;\;R_0>R_c \\2R_0 & \dis \mbox{ for } \; \; R_0<R_c \end{array}\right. =24\left({1\over 2R_0}+R_0\right)-24\left\abs{1\over 2R_0}-R_0\right\abs.
\ee
Since these modes have $\bar a=0$, they are conventionally thermalized (see Eq. (\ref{dressed})). As a result, the deformed canonical ensemble in the hybrid models is nothing but a conventional one for thermal radiation in each phase. The non-analyticity of $Z$ at the fermionic point arises from the appearance of additional massless states in the $O_8\b V_{24}$ sector, which are responsible for an $SU(2)_L$ enhancement. Remarkably, these modes  trigger the winding $\to$ momentum phase transition, as can be seen as follows. For $R_0>R_c$, the Matusbara excitations are pure momentum states in the temporal direction (see Eq. (\ref{ZhybridS})) and satisfy $p_L=p_R\in \ZZ$ or $\ZZ+{1\over 2}$ at the fermionic point. Since the additional massless states have $p_L=\pm 1$, $p_R=0$, the action of their vertex operators $O_\pm$ on the momemtum states can produce pure winding states, $-p_L=p_R$.  Moreover, the presence of a factor $\psi^0_L$ in the expressions of $O_\pm$ implies  the chirality exchange $S_8\leftrightarrow C_8$ as well.  

The effective actions in the winding and momentum phases are just dilaton gravity in two-dimendional Minkowski space, coupled to the free energy associated to thermal radiation. At the transition, the Euclidean scalars $\chi_i$ which are exceptionally massless can have non-trivial backgrounds determined by their tree level equations of motion. In total, the action valid for all times in the case of homogeneous and isotropic evolutions is, when we choose the laps function $N\equiv \beta$,  
\begin{eqnarray}
S&\!\!\!\!=&\!\!\!\!\int dt dx \, \beta a  \left[e^{-2\phi}\left({R\over 2}+2(\partial \phi)^2\right)-{48\pi \over \beta^2}\right]+S_B,\qquad \mbox{with}\nonumber\\
S_B&\!\!\!\!=&\!\!\!\!\int dx \, a(0)\, e^{-2\phi(0)}  \left(-a(0)^{-2}\, {d\chi_i\over dx}{d\b \chi_i\over dx}\right)=-\kappa \int dt dx \, a   \, e^{-2\phi}  \, \delta(t) ,
\end{eqnarray}
where $\kappa\ge 0$ arises from non-trivial gradients of the $\chi_i$'s and can be interpreted as a spacelike brane tension localized at $t=0$. Solving the equations of motion leads to a non-singular bouncing cosmology, 
\be
e^\phi={e^{\phi_c}\over \sqrt{1+ \omega\, \abs t\abs}}\; , \qquad a={a_c\over \sqrt{1+ \omega\, \abs t\abs}}\, e^{{\omega\over 2}\abs t\abs}\; , \qquad T=T_{\rm c} \, e^{-{\omega\over2}\abs t\abs}\, \sqrt{1+ \omega\, \abs t\abs},
\ee
where $a_c$ and $\phi_c$ are arbitrary integrations constants, while $\omega:= {\pi\over \sqrt{2}}\kappa =\sqrt{2} \, e^{\phi_c} \sqrt{48\pi}$ and $T_{\rm c}=\sqrt{2}/(2\pi)$. This evolution is compatible with perturbation theory when the maximal string coupling $e^{\phi_c}$ which occurs at the phase transition is chosen to be sufficiently small. Higher derivative terms are also negligible since the maximal  scalar curvature $R$ is reached at the bounce and is of order $\O(e^{2\phi_c})$. The bounce in the dilaton field is a direct consequence of the negative contribution to the pressure of the brane tension. This is in contrast with other pre- Big Bang scenarios \cite{Gasperini:1996in,Gasperini:2002bn}.


\section*{Acknowledgement}
I am very grateful to the organizers of this conference for the opportunity to present these results. 
This work is partially supported by the contracts PITN-GA-2009-237920,  ERC-AdG-226371, ANR 05-BLAN-NT09-573739,  CEFIPRA/IFCPAR 4104-2, PICS France/Cyprus, France/Greece, France/USA and a PEPS.

\end{document}